\documentclass[a4paper,12pt]{article}

\usepackage{amsmath,amssymb,mathtools} 
\usepackage{color}
\usepackage{graphicx}
\usepackage{subfigure}
\usepackage{cite}           
\usepackage{hyperref}            
\usepackage{multirow,makecell}   
\usepackage{textcomp}
\usepackage{wasysym}
\usepackage{setspace}
\usepackage{verbatim}
\usepackage{float}
\usepackage{ulem}
\usepackage[utf8]{inputenc}

\usepackage[text={17.2cm,25.2cm},centering]{geometry}

\numberwithin{equation}{section}

\begin{document}
	
	\title{\textbf{Phase transition of hot dense QCD Matter from a refined holographic EMD model}}
	\author{Zhibin Li $^{1}$\footnote{lizhibin@zzu.edu.cn},
                Fei Wang $^{1}$\footnote{feiwang@zzu.edu.cn}}
	\date{}
	\maketitle
	
	\vspace{-10mm}
	
	\begin{center}
		{\it
			$^{1}$ Institute for Astrophysics, School of Physics, Zhengzhou University, Zhengzhou 450001, China\\ \vspace{1mm}	
		}
		\vspace{10mm}
	\end{center}

	\begin{abstract}
In this study, we begin by delineating the Einstein-Maxwell-Dilaton (EMD) model within the holographic QCD framework and deriving the equation of state through holographic renormalization. Subsequently, we utilize the Wald method to determine the first law of thermodynamics for a five-dimensional black hole, thereby confirming the alignment of our EMD model with the thermodynamics of the grand canonical ensemble in the boundary field theory. By employing the $n-2$ form in the Wald method, we proceed to calculate the shear viscosity in Gauss-Bonnet gravity. The results we obtain demonstrate consistency with those from first-order metric perturbation. Furthermore, we construct a refined EMD model that attains quantitative agreement with the lattice QCD equation of state and baryon number susceptibility data at finite density. Using this enhanced model, we delve into the investigation of signals related to the QCD phase transition critical endpoint. At present, no non-monotonic changes in $\kappa\sigma^2$ have been observed within the 7$\sim$200 GeV collision energy range at STAR. However, our theoretical analysis suggests that if the chemical freeze-out line does not intersect the first-order phase transition line, a peak-like structure in $\kappa\sigma^2$ is anticipated within the 3$\sim$5 GeV range.
	\end{abstract}
	
	\baselineskip 18pt
	\thispagestyle{empty}
	\newpage
	
	\tableofcontents
	
	\maketitle

\section{Introduction}\label{sec:01}

The phase structure of quark-gluon matter has long been a key focus in heavy ion collision experiments at Relativistic Heavy Ion Collider (RHIC) and the Large Hadron Collider (LHC). Lattice QCD calculations have ruled out the existence of a critical end point (CEP) when $\mu_B/T \leq 3$ and $\mu_B < 300~\text{MeV}$ ~\cite{Vovchenko:2017gkg,Borsanyi:2020fev,Bazavov:2020bjn,Borsanyi:2021sxv,Bollweg:2022fqq,Philipsen:2021qji}. Theoretical studies indicate that the CEP's critical physics likely influences its immediate vicinity. The ratios of $\chi_n^B$ (baryon number susceptibility) exhibit non-monotonic behavior near the CEP ~\cite{Asakawa:2009aj,Schaefer:2011ex,Fan:2016ovc,Portillo:2016fso,Fan:2017mrk,Li:2017ple,Li:2018ygx,Fu:2021oaw,Zhao:2023xpj,Li:2023mpv}. Along the phase boundary, non-monotonic conserved charge fluctuations with $T$ and $\mu_B$ may arise from the CEP' s critical physics, making them potential CEP signals ~\cite{Stephanov:2011pb}. Experimental results suggest that for $\mu_B/T \leq 3$, there is a significant overlap between the chemical freeze-out region and the crossover region ~\cite{STAR:2017sal}. If the freeze-out line passes near the CEP, similar non-monotonicity might be observed in $\chi_n^B$ and their ratios along the freeze-out line ~\cite{Fan:2016ovc,Fan:2017mrk,Li:2017ple,Portillo:2017gfk,Li:2018ygx,Wang:2018sur,Fu:2021oaw,Zhao:2023xpj,Huang:2023ogw}.

Notably, the ratios of $\chi_n$, including $\chi_n^B$, $\chi_n^Q$, and $\chi_n^S$, directly correlate with measurable quantities such as the variance $\sigma^2$ and kurtosis $\kappa$ of conserved charge distributions. This correlation allows for locating the CEP through experimental measurements of these susceptibilities. However, the latest STAR data on net proton number fluctuations shows no clear evidence of a CEP at collision energies $\sqrt{s_{NN}} >$ 7.7 GeV \cite{STAR:2025zdq}. Within the 7.7 to 200 GeV collision energy range, the ratio of the fourth-order to the second-order net proton number susceptibility, $\kappa\sigma^2$, exhibits a downward trend as collision energy decreases without significant non-monotonic behavior \cite{STAR:2025zdq}. This highlights the need for more precise CEP location information and new theoretical support.

In recent decades, the holographic approach based on gauge/gravity duality has been widely applied to the study of QCD matter properties. The Einstein-Maxwell-Dilaton system, introducing a nonconformal bulk dilaton scalar and a $U(1)$ gauge field, has been widely used in holographic QCD research \emph{e.g.}~\cite{ DeWolfe:2010he, DeWolfe:2011ts, Cai:2012xh, Cai:2012eh, Finazzo:2013efa, Yang:2014bqa, Critelli:2017oub, Li:2017ple, Chen:2017cyc, Knaute:2017opk, Fang:2018axm, Ballon-Bayona:2020xls, Li:2020hau, Grefa:2021qvt, He:2022amv, Grefa:2022fpu,Chen:2024ckb}. The advancement in numerical calculation methods, particularly the application of machine learning, has enabled these studies to achieve significant progress in numerical consistency with lattice QCD and has offered a substantial theoretical foundation for determining the CEP location \cite{Cai:2022omk,Hippert:2023bel,Chen:2024ckb,Luo:2024iwf,Zhu:2025gxo,Sun:2025uga}. 

Current EMD models for holographic QCD, typically Gubser based, mainly fall into two categories. One category involves inputting the dilaton potential and gauge kinetic function, numerically solving the equations of motion to derive the equation of state \cite{DeWolfe:2010he, DeWolfe:2011ts, Finazzo:2013efa, Critelli:2017oub, Knaute:2017opk, Grefa:2021qvt, He:2022amv, Grefa:2022fpu, Li:2024lrh}. The other category employs a potential reconstruction method by inputting the warp factor and gauge kinetic function or the dilaton field, which can yield integral form solutions to the equations of motion \cite{Cai:2012xh, Cai:2012eh, Yang:2014bqa, Li:2017ple, Chen:2017cyc, Fang:2018axm, Ballon-Bayona:2020xls, Li:2020hau, Chen:2024ckb, Li:2025ugv}. When obtaining thermodynamic quantities, methods like directly reading temperature, entropy density, chemical potential, and particle number density, then integrating thermodynamic relations for pressure and energy density are available \cite{DeWolfe:2010he, DeWolfe:2011ts}. Alternatively, holographic renormalization of the 5-dimensional EMD action enables direct acquisition of required thermodynamic quantities  \cite{Cai:2022omk, Li:2023mpv}. The thermodynamic relations of the above EMD models are also worth studying. Generally, the thermodynamic relations of the boundary can be derived through black hole thermodynamic laws. Follow the Wald method \cite{Wald:1993nt,Iyer:1994ys} the gravity with dilaton field has been studied in \cite{Liu:2013gja,Lu:2014maa,Liu:2017kml,Li:2020spf}. In this work, we follow the similar approach to demonstrate the first law of thermodynamics for the EMD model. 

As QCD matter near phase boundaries behaves as a strongly coupled fluid \cite{Shuryak:2014zxa}, its transport properties are crucial for studying QCD matter characteristics. In EMD models, the shear viscosity to entropy density ratio is constant as $\eta/s=1/4\pi$ \cite{DeWolfe:2011ts}. To go beyond this constant, higher derivative gravity like the Gauss-Bonnet term can be introduced \cite{Ge:2009eh,Charmousis:2012dw}. In \cite{Fan:2018qnt} suitable metric perturbations allow calculating the shear viscosity to entropy density ratio under higher derivative gravity. In this work, we analytically derive the behavior of the shear viscosity to entropy density ratio in both Einstein gravity and Gauss-Bonnet gravity.

In the Gubser model and subsequent holographic EMD models \cite{DeWolfe:2010he, DeWolfe:2011ts, Cai:2022omk, Rougemont:2023gfz, Li:2024lrh}, setting the UV limit of the gauge kinetic function to 1 can lead to $\chi_2^B$ values at high temperatures that far exceed the Stefan–Boltzmann limit (SB limit) \cite{Fu:2020xnh, Jokela:2024xgz}. To address this, we treat the UV limit of the gauge kinetic function as an undetermined parameter \cite{Jokela:2024xgz} and determine the model parameters by fitting HotQCD equation of state and baryon number susceptibility data at zero density \cite{HotQCD:2012fhj, HotQCD:2014kol, Bazavov:2017dus}. We then compare model results with HotQCD data on various thermodynamic quantities and baryon number susceptibilities at finite density \cite{Bazavov:2017dus,Bollweg:2022rps}. Next, we compare the model’s baryon number susceptibility ratios along the chemical freeze-out line with experimental results for net proton number susceptibility ratios \cite{Xu:2016mqs,STAR:2021fge,STAR:2025zdq}. Finally, by fitting different chemical freeze-out lines, we identify potential CEP signals.

The rest of this paper is organized as follows. In section~\ref{sec:02}, we detail the EMD model's setup and holographic renormalization. In section~\ref{sec:03}, we use the Wald method to demonstrate the first law of thermodynamics for the EMD model. In section~\ref{sec:04}, we analytically derive shear viscosity in Gauss - Bonnet gravity. In section~\ref{sec:05}, we present a refined EMD model for quantitative analysis of QCD matter phase transitions. We conclude with some discussion in Section~\ref{sec:06}.

\section{The Einstein-Maxwell-dilaton model}\label{sec:02}
\subsection{Basic settings of EMD model}
The action of Einstein-Maxwell-dilaton model is
\begin{eqnarray}
    S=\frac{1}{2\kappa_5^2}\int d^5x \sqrt{-g}\mathcal{L}=\frac{1}{2\kappa_5^2}\int d^5x \sqrt{-g}\left[\mathcal{R}-\frac{1}{2}\nabla_{\mu}\phi\nabla^{\mu}\phi-V(\phi)-\frac{Z(\phi)}{4}F^{\mu\nu}F_{\mu\nu}\right] .
\end{eqnarray}
And the equations of motion are
\begin{equation}
    \nabla_{\mu}\nabla^{\mu}\phi-\frac{Z'(\phi)}{4}F^{\mu\nu} F_{\mu\nu}-V'(\phi)=0,~~~
    \nabla^\rho \left[Z(\phi) F_{\rho\lambda}\right]=0, \nonumber
\end{equation}
\begin{equation}
    \mathcal{R}_{\mu\nu}-\frac{1}{2}g_{\mu\nu}\mathcal{R}=\frac{1}{2}\nabla_{\mu}\phi\nabla_{\nu}\phi+\frac{Z(\phi)}{2}F_{\mu\rho}F_{\nu}{}^{\rho}+\frac{1}{2} g_{\mu\nu} \left[-\frac{1}{2}\nabla_{\rho}\phi\nabla^{\rho}\phi-V(\phi)-\frac{Z(\phi)}{4}F^{\lambda\sigma}F_{\lambda\sigma}  \right].
\end{equation}
The metric and field take the following form
\begin{equation}
    ds^2=-e^{-w(r)}f(r)dt^2+\frac{1}{f(r)}dr^2+r^2d\vec{x}^2 , \nonumber
\end{equation}
\begin{equation}
    \phi=\phi(r), ~~\text{and}~~A_{\mu}dx^{\mu}=A_t(r)dt,
\end{equation}
where $d\vec{x}^2=dx^2+dy^2+dz^2$. Then the equations of motion are
\begin{eqnarray}
    w'+\frac{1}{3}r \phi'^2=0, \nonumber \\ 
    \phi''+\left( \frac{3}{r}+\frac{f'}{f}-\frac{w'}{2} \right)\phi'+\frac{1}{2f}Z'(\phi) e^{w}  A_t'^2-\frac{1}{f}V'(\phi)=0, \nonumber \\
    \partial_r \left[ e^{w/2}r^3 Z(\phi) A_t' \right]=0, \nonumber \\
    \frac{4}{r^2}+\frac{2V(\phi)}{3f}+\frac{Z(\phi)e^{w}A_t'^2}{3f}+\frac{2f'}{rf}-\frac{w'}{r}=0.
\end{eqnarray}
By setting the cosmological constant $\Lambda=-6$ and the dimension of $\phi$ to 3, we obtain the UV asymptotic behavior of $V(\phi)$ as
\begin{equation}
    V(\phi\rightarrow 0)=-12-\frac{3}{2}\phi^2+\mathcal{O}(\phi^3).
\end{equation}
And the UV expansions are
\begin{eqnarray}
   \left. f\right| _{r\rightarrow \infty}&=&r^2+\frac{\phi_s^2}{6}-\frac{2\phi_s^3V^{(3)}(0)}{9r}+\frac{f_v}{r^2}+\frac{\phi_s^4}{24r^2}\left[3V^{(3)}(0)^2-V^{(4)}(0)-2\right] \ln(r) +\mathcal{O}(r^{-3}), \nonumber \\
   \left.   w\right| _{r\rightarrow \infty}&=&\frac{\phi_s^2}{gr^2}-\frac{2\phi_s^3 V^{(3)}(0)}{9r^3}+\mathcal{O}(r^{-4}), \nonumber  \\
    \left.  \phi\right| _{r\rightarrow \infty}&=&\frac{\phi_s}{r}-\frac{\phi_s^2 V^{(3)}(0)}{2r^2}+\frac{\phi_v}{r^3}+\frac{\phi_s^3}{12 r^3}\left[3V^{(3)}(0)^2-V^{(4)}(0)-2 \right]\ln(r) +\mathcal{O}(r^{-4}), \nonumber  \\
    \left.  A_t\right| _{r\rightarrow \infty}&=& \mu_B-\frac{a_v}{2r^2}+\frac{a_v \phi_s Z'(0)}{3Z(0) r^3}+\mathcal{O}(r^{-4}),
\end{eqnarray}
with $\phi_s$, $\phi_v$, $f_v$ and $a_v$ the coefficients should be fixed by the equation of motion. And $\mu_B$ the baryon chemical potential. By imposing regular boundary conditions at the IR boundary, we can derive
\begin{eqnarray}
    \left. f\right| _{r\rightarrow r_h}&=&-\frac{r_h}{6}\left[ 2V(\phi_h)+A_h^2 Z(\phi_h) \right](r-r_h)+\mathcal{O}((r-r_h)^{2}) ,\nonumber \\
    \left.   w\right| _{r\rightarrow r_h}&=& -\frac{3}{r_h}\left[\frac{-2V'(\phi_h)+A_h^2Z'(\phi_h)}{2V(\phi_h)+A_h^2Z(\phi_h)}\right]^2(r-r_h) +\mathcal{O}((r-r_h)^{2}) ,\nonumber \\
    \left.  \phi\right| _{r\rightarrow r_h}&=&\phi_h-\frac{6V'(\phi_h)-3A_h^2 Z'(\phi_h)}{2r_h V(\phi_h)+A_h^2 r_h Z(\phi_h)}(r-r_h)+\mathcal{O}((r-r_h)^{2}), \nonumber \\
   \left.  A_t\right| _{r\rightarrow r_h}&=& A_h(r-r_h)+\mathcal{O}((r-r_h)^{2}),
\end{eqnarray}
with $r_h$, $\phi_h$ and $A_h$ the input coefficients.
Then the Hawking temperature and black hole entropy can be calculated as
\begin{eqnarray}
    T&=&\frac{1}{4\pi}\sqrt{e^{-w(r_h)}f'(r_h)\left[f'(r_h)-f(r_h)w'(r_h)\right]}=\frac{1}{4\pi} f'(r_h),\\
    S_{BH}&=&\frac{2\pi}{\kappa_5^2}\int_{r=r_h}\sqrt{\tilde{h}}dx^1dx^2dx^3=\frac{2\pi}{\kappa_5^2}\tilde{V}r_h^3,
\end{eqnarray}
where $\tilde{h}$ is the determinant of the induced metric on the horizon as $\tilde{h}_{ab}dx^adx^b=r^2 d\vec{x}^2$. And $\tilde{V}$ is the volume of the total space. And the entropy density
\begin{equation}
    s=\frac{2\pi}{\kappa_5^2}r_h^3.
\end{equation}

\subsection{Holographic renormalization}
To calculate the free energy density, we compute the on shell action as
\begin{eqnarray}
     S_{on-shell}&=&\left. -\frac{1}{\kappa_5^2}\frac{\tilde{V}}{T} e^{-\frac{w}{2}}r^2f\right|_{r_h}^{\infty} \nonumber  \\
    &=& \frac{1}{2\kappa_5^2}\frac{\tilde{V}}{T}\left\{-2r^4-\frac{\phi_s^2}{6}r^2+\frac{2}{9}\phi_s^3 V^{(3)}(0)r+\frac{1}{24}\phi_s^4\left[V^{(4)}(0)-3V^{(3)}(0)^2+2\right]\ln(r) \right.\nonumber \\
    &&\left.\left. -2f_v+\frac{\phi_s^4}{36}+\frac{1}{2}\phi_s\phi_v+\frac{7}{96}\phi_s^4V^{(3)}(0)^2+\frac{1}{288}\phi_s^4V^{(4)}(0)\right\}\right|_{r\rightarrow \infty}
\end{eqnarray}
which is divergent. To renormalize the UV divergence we use the induced metric on the hypersurface with fixed $r$ $h_{\mu\nu}=g_{\mu\nu}-n_{\mu}n_{\nu}$ with $n=\partial_r$ the outward normal vector. And the extrinsic curvature
\begin{eqnarray}
    K_{\mu\nu}=-\frac{1}{2}\left(\nabla_\mu n_\nu+\nabla_\nu n_\mu \right),~~K=h^{\mu\nu}K_{\mu\nu}=\frac{rf'+f\left(6-rw'\right)}{2r\sqrt{f}}.
\end{eqnarray}
The contributions of different terms to the renormalization are as follows
\begin{eqnarray}
   && \hspace{-0.8cm} \left. \sqrt{-h}K\right|_{r\rightarrow\infty}=4r^4+\frac{\phi_s^2}{3}r^2-\frac{4}{9}\phi_s^3 V^{(3)}(0)r-\left[\frac{\phi_s^4}{6}-\frac{1}{4}\phi_s^4V^{(3)}(0)^2 +\frac{1}{12}\phi_s^4v^{(4)}(0)\right]\ln(r)+2f_v-\frac{\phi_s^4}{72}, \nonumber \\
   && \hspace{-0.4cm}\left. \sqrt{-h}\right|_{r\rightarrow\infty}= r^4+\frac{f_v}{2}-\frac{1}{4}\phi_s\phi_v-\frac{1}{576}\phi_s^4\left[6+21 V^{(3)}(0)+V^{(4)}(0)\right], \nonumber \\
   && \hspace{-0.8cm}\left. \sqrt{-h}\phi^2\right|_{r\rightarrow\infty}= \phi_s^2r^2-\phi_s^3V^{(3)}(0)r-\frac{\phi_s^4}{6}\left[2-3V^{(3)}(0)^2 +V^{(4)}(0)\right]\ln(r)  +2\phi_s\phi_v+\frac{1}{4}\phi_s^4V^{(3)}(0)^2 ,\nonumber \\
   && \hspace{-0.8cm}\left. \sqrt{-h}\phi^3\right|_{r\rightarrow\infty}= \phi_s^3 r -\frac{3}{2}\phi_s^4V^{(3)}(0), \nonumber \\
   && \hspace{-0.8cm} \left. \sqrt{-h}\phi^4\right|_{r\rightarrow\infty}= \phi_s^4 ,\nonumber \\
  && \hspace{-0.8cm} \left. \sqrt{-h}F_{\mu\nu}F^{\mu\nu}\right|_{r\rightarrow\infty}= 0.
\end{eqnarray}

The boundary term can be fixed as
\begin{eqnarray}
  S_{\partial}&=&\frac{1}{2\kappa_5^2}\int d^4x \sqrt{-h} \mathcal{L}_{\partial} \nonumber   \\
  &=&\frac{1}{2\kappa_5^2}\int d^4x \sqrt{-h}\left.\left\{2K-6-\frac{1}{2}\phi^2 +\frac{1}{6}V^{(3)}(0)\phi^3-b\phi^4\right.\right. \nonumber  \\
  && \left.\left.+\frac{1}{24}\left[ 2-3V^{(3)}(0)^2+V^{(4)}(0) \right]\phi^4\ln(r)
   \right\}\right|_{r=\infty} 
\end{eqnarray}
And the renormalzed on shell action
\begin{equation}
    \frac{T}{\tilde{V}}\left(S+S_{\partial}\right)_{on-shell}=f_v-\phi_s\phi_v-\frac{\phi_s^4}{144}\left[ 9-144b-12 V^{(3)}(0)^2+2V^{(4)}(0) \right].
\end{equation}
The metric on the boundary $\hat{h}_{ab}=\left. h_{ab}\right|_{r=\infty}=r^2  \zeta_{ab} $ where $\zeta_{ab}$ is the metric of Minkowski spacetime. Then the energy momentum tensor on the boundary is
\begin{eqnarray}
 T_{ab}&=&\frac{2}{\sqrt{-\zeta}}\frac{\delta \left(S+S_{\partial}\right)_{on-shell}}{\delta \zeta^{ab}}=\lim_{r\rightarrow\infty}\left(\frac{2r^2}{\sqrt{-h}}\frac{\delta \left(S+S_{\partial}\right)_{on-shell}}{\delta h^{ab}} \right) \nonumber  \\
 &=&\frac{1}{2\kappa_5^2}\lim_{r\rightarrow \infty} r^2\left\{ -2K_{ab}-Z(0)F_{ac}F_{b}{ }^{c}\ln(r) +h_{ab}\mathcal{L}_{\partial} \right\}.
\end{eqnarray}
The presure and energy density can be calculated as
\begin{eqnarray}
    \mathcal{E}&=&T_{00}=\frac{1}{2\kappa_5^2}\left\{-3f_v+\phi_s\phi_v+\phi_s^4\left[ \frac{1}{48}+b+\frac{3}{8}V^{(3)}(0)^2 \right]\right\},\\
    P_x&=&P_y=P_z=T_{ii} \nonumber \\
    &=&\frac{1}{2\kappa_5^2}\left\{-f_v+\phi_s\phi_v+\frac{\phi_s^4}{144}\left[ 9-144b-12 V^{(3)}(0)^2+2V^{(4)}(0)\right]\right\}\\
    &=&- \frac{T}{\tilde{V}}\left(S+S_{\partial}\right)_{on-shell} ,\nonumber \\
    I&=&\mathcal{E}-3P=-\frac{1}{2\kappa_5^2}\left\{2\phi_s\phi_v+\frac{\phi_s^4}{24}\left[ 4-96b-15 V^{(3)}(0)^2+V^{(4)}(0)\right]\right\}.
\end{eqnarray}
For the sources $\phi_s$ we have the condensates
\begin{eqnarray}
    \langle \Phi\rangle &=&\frac{1}{2\kappa_5^2} \lim_{r\rightarrow \infty}\frac{\sqrt{-h}}{r}\left\{- n_\mu\nabla^\mu\phi -\phi+\frac{1}{2}V^{(3)}(0)\phi^2-4b\phi^3+\frac{1}{6}\left[ 2-3 V^{(3)}(0)^2+V^{(4)}(0) \right]\phi^3\ln(r) \right\} \nonumber  \\
    &=&\frac{1}{2\kappa_5^2}\left\{ 2\phi_v+\frac{1}{12}\phi_s^3\left[3-48b-9V^{(3)}(0)^2+V^{(4)}(0)\right]\right\}.
\end{eqnarray}

\section{The first law of thermodynamics}\label{sec:03}
Follow the Wald method \cite{Wald:1993nt,Iyer:1994ys} the gravity with dilaton field has been studied in \cite{Liu:2013gja,Lu:2014maa,Liu:2017kml,Li:2020spf}.
In order to provide the first law of thermodynamics for black holes in EMD model, we choose the vector $\xi=\partial_t$ which is a time-like Killing vector. Then we have 
\begin{equation}
    \delta H_{\xi}=\int_{\partial\mathcal{C}}\left(\delta\mathbf{Q}-i_\xi \mathbf{\Theta}_\delta\right)=0.
\end{equation}
For our case the bulk $\mathcal{C}$ is the entire space, with its boundaries being the black hole horizon $r=r_h$ and the infinite boundary $r=\infty$. Both $\mathbf{j}$ and $\mathbf{\Theta}$ are only related to $r$, so the integration of three-dimensional space is trivial, we have
 \begin{equation}
    \delta H_{\xi}=\int_{\partial\mathcal{C}}\left(\delta\mathbf{Q}-i_\xi \mathbf{\Theta}_\delta\right)=\delta H_{\xi}^{\infty}-\delta H_{\xi}^{h}=0.
\end{equation}
We firstly calculate the $n-2$ form $\mathbf{Q}$. For $\xi=\partial_t$ we have \cite{Fan:2014ala,Chen:2016qks,Fan:2018qnt}
\begin{eqnarray}
    \mathbf{Q}_{g}&=&\frac{1}{3!}\epsilon_{\mu\nu\alpha\beta\gamma}\sqrt{-g}\left(2\frac{\partial\mathcal{L}}{\partial R_{\mu\nu\rho\sigma}}\nabla_\rho\xi_\sigma+4\xi_\rho\nabla_\sigma \frac{\partial\mathcal{L}}{\partial R_{\mu\nu\rho\sigma}}\right)  dx^\alpha \wedge dx^\beta \wedge dx^\gamma    \nonumber \\
    &=& e^{-\frac{w}{2}}r^3\left(f'-fw' \right)  dx \wedge dy \wedge dz,\\
    \mathbf{Q}_{A}&=&\frac{1}{3!}\epsilon_{\mu\nu\alpha\beta\gamma}\sqrt{-g}Z\left(\xi^\rho A_\rho F^{\mu\nu}\right)  dx^\alpha \wedge dx^\beta \wedge dx^\gamma  \nonumber  \\
    &=& -e^{-\frac{w}{2}}r^3 e^w A_t Z A_t'  dx \wedge dy \wedge dz,\\
     \mathbf{Q}_{\phi}&=& 0.
\end{eqnarray}
where $\epsilon_{\mu\nu\alpha\beta\gamma}$ is the Levi-Civita tensor with $\epsilon_{trxyz}=1$. And 
\begin{equation}
     \mathbf{Q}=\mathbf{Q}_{gr}+\mathbf{Q}_{A}+\mathbf{Q}_{\phi}=-e^{-\frac{w}{2}}r^3 \left(-f'+fw' +e^w A_t Z A_t'\right)  dx \wedge dy \wedge dz.
\end{equation}
Then we consider the variation $\Psi \rightarrow \Psi+\delta \Psi$ and $\delta \left(\Psi'\right)=\left(\delta\Psi\right)'$ then we have  
\begin{equation}
    \delta \mathbf{Q}=\delta\left[e^{-\frac{w}{2}}r^3 \left(  f'-fw' \right)   \right]-\delta\left( e^{\frac{w}{2}}r^3A_t Z A_t'\right),
\end{equation}
and
\begin{eqnarray}
    \mathbf{\Theta}_{g}&=&  \frac{2}{4!}\epsilon_{\mu\nu\rho\sigma\tau}\sqrt{-g}\left(\frac{\partial\mathcal{L}}{\partial R_{\alpha\beta\gamma\mu}}\nabla_\alpha \delta g_{\beta\gamma}-\nabla_\alpha \frac{\partial\mathcal{L}}{\partial R_{\alpha\beta\gamma\mu}}\delta g_{\beta\gamma}\right) dx^\nu \wedge dx^\rho \wedge dx^\sigma \wedge dx^\tau  \nonumber \\
    &=& \left\{\delta \left[e^{-\frac{w}{2}}r^3 \left(  f'-fw' \right)   \right]+3e^{-\frac{w}{2}}r^2\delta f \right\}dt\wedge dx\wedge dy\wedge dz ,  \\
    \mathbf{\Theta}_{\phi}&=&  -\frac{1}{4!}\epsilon_{\mu\nu\rho\sigma\tau}\sqrt{-g} \nabla^\mu\phi \delta\phi dx^\nu \wedge dx^\rho \wedge dx^\sigma \wedge dx^\tau \nonumber \\
    &=& e^{-\frac{w}{2}}r^3f\phi'\delta\phi dt\wedge dx\wedge dy\wedge dz , \\
    \mathbf{\Theta}_{A}&=& \frac{1}{4!}\epsilon_{\mu\nu\rho\sigma\tau}\sqrt{-g} Z F^{\mu\alpha}\delta A_\alpha dx^\nu \wedge dx^\rho \wedge dx^\sigma \wedge dx^\tau \nonumber \\
    &=&  - e^{\frac{w}{2}}r^3Z A_t' \delta A_t dt\wedge dx\wedge dy\wedge dz.
\end{eqnarray}
Then we have
\begin{eqnarray}
\mathbf{\Theta}&=&\mathbf{\Theta}_{g}+\mathbf{\Theta}_{\phi}+\mathbf{\Theta}_{A} \nonumber \\
    &=&\left\{\delta \left[e^{-\frac{w}{2}}r^3 \left(  f'-fw' \right)   \right]+3e^{-\frac{w}{2}}r^2\delta f+e^{-\frac{w}{2}}r^3f\phi'\delta\phi- e^{\frac{w}{2}}r^3Z A_t' \delta A_t \right\}  \nonumber \\
    && \hspace{5mm} dt\wedge dx\wedge dy\wedge dz.
\end{eqnarray}
Finally, we get 
\begin{eqnarray}
    &&\delta \mathbf{Q}-i_\xi \mathbf{\Theta} =-\left[3 e^{-\frac{w}{2}}r^2\delta f+\frac{1}{2}e^{\frac{w}{2}}r^3A_t Z A_t'\delta w +e^{-\frac{w}{2}}r^3 \left( e^w A_t Z'A_t'+f\phi' \right) \delta \phi+e^{\frac{w}{2}}r^3A_t Z\delta A_t'\right] \nonumber \\
    && \hspace{4cm} dx\wedge dy\wedge dz.
\end{eqnarray}
Here we assume that the variation comes from changes in parameters $\phi_s$, $\phi_v$, $f_v$, $a_v$, $r_h$, $A_h$ and $\phi_h$, while the coordinates remain unchanged. At the UV boundary we have
\begin{eqnarray}
    \delta H_{\xi}^{\infty}&=&\tilde{V}\left\{\left[3\phi_v+\frac{1}{12}\phi_s^3\left(4+9 V^{(3)}(0)^2+V^{(4)}(0)\right)\right]\delta \phi_s+\phi_s\delta\phi_v-3\delta f_v-\mu_B Z(0) \delta a_v\right\} \nonumber \\
    &=&\tilde{V}\left[2\kappa_5^2 \delta \mathcal{E}-\mu_B Z(0) \delta a_v+2\kappa_5^2\langle \Phi\rangle\delta \phi_s\right].
\end{eqnarray}
According to the asymptotic expansion of each field at the IR boundary $A_t(r_h)=f(r_h)=w(r_h)=0$, it can be concluded as
\begin{eqnarray}
\delta H_{\xi}^{h}=-\left.3\tilde{V} e^{-\frac{w}{2}}r^2\delta f\right|_{r=r_h}=3 \tilde{V}r_h^2 f'(r_h)\delta r_h=2\kappa_5^2\tilde{V} T\delta s.
\end{eqnarray}
Finally, we get the relation
\begin{equation}
    \delta H_{\xi}^{\infty}=\delta H_{\xi}^{h} =2\kappa_5^2 \delta \mathcal{E}-\mu_B Z(0) \delta a_v+2\kappa_5^2\langle \Phi\rangle\delta \phi_s=2\kappa_5^2 T\delta s.
\end{equation}
We can define the baryon density 
\begin{equation}
    \rho_B=\frac{Z(0) }{2\kappa_5^2}a_v.
\end{equation}
Then we have the thermodynamic relation
\begin{equation}
     \delta \mathcal{E}=\mu_B \delta \rho_B+ T\delta s-\langle \Phi\rangle\delta \phi_s.
\end{equation}
Once we fix $\phi_s$ to a constant for different temperature and baryon chemical potential then we get the first law of thermodynamics for black hole
\begin{eqnarray}\label{eq418}
    \delta \mathcal{E}=\mu_B \delta \rho_B+ T\delta s
\end{eqnarray}
In fact, $\mathbf{Q}$ is similar to the on shell action $S_{on-shell}$ for $\xi=\partial_t$ because
\begin{equation}
    \mathbf{j}=-i_{\xi}L=-\sqrt{-g}\mathcal{L}dr \wedge dx \wedge dy \wedge dz=d\mathbf{Q}.
\end{equation}
As $-\sqrt{-g}\mathcal{L}$ is a function of $r$ then we have
\begin{equation}
    \mathbf{Q}=-dx \wedge dy \wedge dz\int_{r_0}^r d\tilde{r} \sqrt{-g(\tilde{r})}\mathcal{L}(\tilde{r}).
\end{equation}
So we have the relation
\begin{equation}
    \frac{\partial}{\partial r}\left(-2 e^{-\frac{w}{2}}r^2f\right)=\frac{\partial}{\partial r}\left[e^{-\frac{w}{2}}r^3 \left(-f'+fw' +e^w A_t Z A_t'\right)\right]=e^{-\frac{w}{2}}r\left[-2rf'+\left(rw'-4\right)f \right].
\end{equation}
Here we simplify the last equation using the equation of motion. Then we get a radial conservation quantity
\begin{eqnarray}
    \mathbf{J}=-2 e^{-\frac{w}{2}}r^2f-e^{-\frac{w}{2}}r^3 \left(-f'+fw' +e^w A_t Z A_t'\right)=e^{\frac{w}{2}}r^3\left[r^2\left(e^{-w}\frac{f}{r^2}\right)' -ZA_tA_t'\right],
\end{eqnarray}
with 
\begin{equation} \label{eq423}
    \frac{\partial}{\partial r}\mathbf{J}=0.
\end{equation}
At the UV and IR boundaries we have
\begin{eqnarray}
    \mathbf{J}_{h}&=&f'(r_h)r_h^3 =2\kappa_5^2 Ts\\
    \mathbf{J}_{\infty}&=& -4f_v+2\phi_s\phi_v+\frac{1}{72}\phi_s^4\left[6+21V^{(3)}(0)^2+V^{(4)}(0)\right]=2\kappa_5^2\left(\mathcal{E}+P\right)-Z[0]\mu_B a_v
\end{eqnarray}
So with the \ref{eq423} we have
\begin{equation}
    \mathbf{J}_{h}-\mathbf{J}_{\infty}=2\kappa_5^2 Ts-2\kappa_5^2  \left(\mathcal{E}+P-\mu_B \rho_B\right)=0
\end{equation}
then we have the thermodynamic relation 
\begin{eqnarray}
    \mathcal{E}+P-\mu_B \rho_B-Ts=0~~~\text{or}~~~P=-\mathcal{E}+\mu_B \rho_B+Ts,
\end{eqnarray}
and with \ref{eq418} we have the relation
\begin{equation}
    \delta P=\delta\left( \mathcal{-E}+\mu_B \rho_B+Ts\right)=\rho_B\delta\mu_B+s\delta T.
\end{equation}

\section{Shear viscosity in Gauss-Bonnet gravity}\label{sec:04}

To calculate the shear viscosity we assume the flow is moving along $y$ axis with the velocity $u=u(x)$. Then we perturb the background as
\begin{equation}
    \delta g_{xx}=j(r) \vartheta t ,~~~\delta g_{xy}=\delta g_{yx}=p(r),
\end{equation}
where $\vartheta=\partial u/\partial x$. And we take the in-going boundary condition \cite{Fan:2018qnt}
\begin{equation}
   \left. p\right|_{r\rightarrow r_h}=\frac{\vartheta r_h^2}{4\pi T}\ln(r-r_h)+\mathcal{O}(( r-r_h )^0).
\end{equation}
The shear viscosity can be calculated by $\mathbf{Q}$ as \cite{Fan:2018qnt}
\begin{equation}
    \eta = -\frac{1}{2\kappa_5^2\vartheta}\sqrt{-g} \mathbf{Q}^{ry}.
\end{equation}
Note that here the Killing vector should be $\xi=\partial_x$. Then for Einstein gravity we have 
\begin{equation}
    \mathbf{Q}^{ry}_{E}= \frac{f}{r^3}\left( 2p-rp' \right) .
\end{equation}
At the IR boundary we have
\begin{equation}
    \eta=\frac{s}{4\pi}.
\end{equation}
And for Gauss-Bonnet gravity 
\begin{equation}
   \hspace{-0.6cm} S=\frac{1}{2\kappa_5^2}\int d^5x \sqrt{-g}\left[\mathcal{R}-\frac{1}{2}\nabla_{\mu}\phi\nabla^{\mu}\phi-V(\phi)-\frac{Z(\phi)}{4}F^{\mu\nu}F_{\mu\nu}+\frac{\lambda_{GB}}{2} \left( \mathcal{R}^2-4\mathcal{R}_{\mu\nu}^2+\mathcal{R}_{\mu\nu\rho\sigma}^2 \right)\right],
\end{equation}
with \cite{Fan:2014ala,Chen:2016qks,Fan:2018qnt}
\begin{equation}
    \mathbf{Q}^{\mu\nu}=-2\left( \nabla^{\left[\mu\right.}\xi^{\left.\nu\right]} +\lambda_{GB}\left( \mathcal{R} \nabla^{\left[\mu\right.}\xi^{\left.\nu\right]}  -4\mathcal{R}^{\sigma \left[ \mu \right.}\nabla_{\sigma}\xi^{\left.\nu\right]}  +\mathcal{R}^{\mu\nu\rho\sigma} \nabla_\rho \xi_\sigma\right) \right),
\end{equation}
and
\begin{equation}
    \mathbf{Q}_{GB}^{ry}=\frac{f}{r^3}\left( 2p-rp' \right) \left( 1-\frac{\lambda_{GB}}{r}f'+\frac{\lambda_{GB}}{r} fw' \right)
\end{equation}.
Then we get
\begin{equation}
    \frac{\eta}{s}=\left( \frac{1}{4\pi}-\frac{\lambda_{GB}T}{r_h}  \right),
\end{equation}
which is consistent with the result in \cite{Charmousis:2012dw}.
For AdS-Schwarzschild black hole $T=\frac{d r_h}{4\pi}$. And for $d=4$ we have
\begin{equation}
    \frac{\eta}{s}=\frac{1}{4\pi}\left( 1-4\lambda_{GB}  \right),
\end{equation}
which aligns with the results in \cite{Brigante:2007nu}. 

\section{A refined EMD model}\label{sec:05}
In this work we set 
\begin{equation}\label{eq51}
\begin{split}
 V(\phi)&=-12 \cosh\left[c_1\phi\right]+\left(6c_1^2-\frac{3}{2}\right)\phi^2+c_2\phi^6 +c_3 \phi^8   \,, \\
Z(\phi)&=c_4 \text{sech}\left[ c_5 \phi^3 \right]\,,
\end{split}
\end{equation}
Compared to the $V(\phi)$ used in \cite{Cai:2022omk,Li:2023mpv}, we added a high order term $c_3 \phi^8$ to optimize the equation of state behavior. This results in better quantitative agreement with lattice QCD results. From the previous calculation in section \ref{sec:03}, we know that $Z(0) = 1$ isn't required. Therefore, in this work, we treat $Z(0)$ as an independent parameter and choose a new $Z(\phi)$. This choice is based on two reasons: First, at extremely high temperatures, the original $Z(\phi)$ causes $\chi_2^B$ to significantly exceed the SB limit \cite{Jokela:2024xgz,Fu:2024wkn}, while the new $Z(\phi)$ avoids this problem. Second, the original $Z(\phi)$ had a spike near $\phi = 0$, where $|Z'(\phi)|$ was very large, potentially causing numerical instability.

\begin{figure}[htbp]
\centering
\includegraphics[width=.43\textwidth]{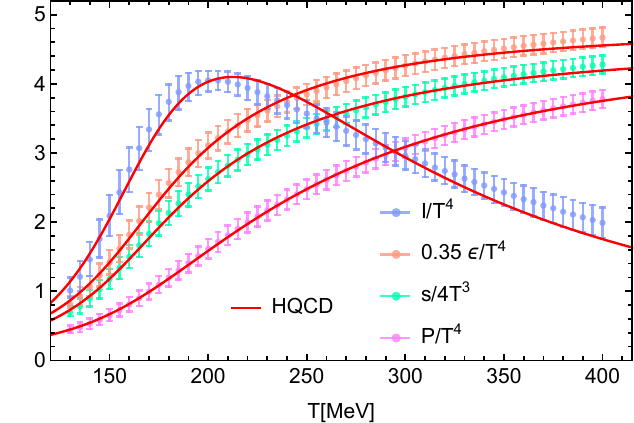}
\qquad
\includegraphics[width=.45\textwidth]{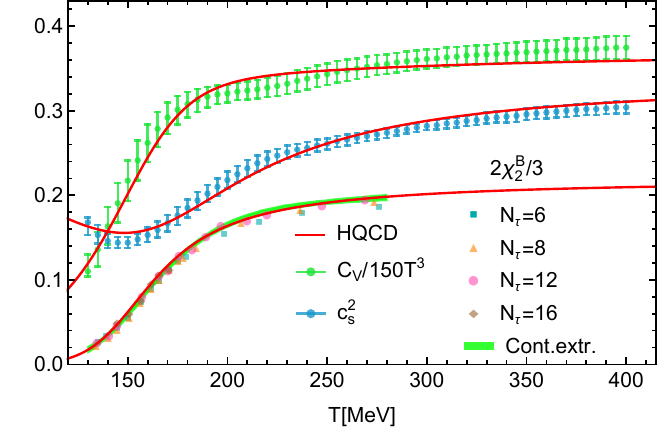}
\caption{A comparison of the equation of state from our holographic model (red solid curves) with the corresponding lattice QCD data at $\mu_B=0$~\cite{HotQCD:2012fhj, HotQCD:2014kol, Bazavov:2017dus}. (Left panel) the entropy density $s/T^3$, the pressure $P/T^4$, the energy density $\epsilon/T^4$, and the trace anomaly $I/T^4$. (Right panel) the specific heat $C_V$, the squared speed of sound $c_s^2$, and the second order baryon number susceptibility $\chi^B_2$.}\label{fig1}
\end{figure}
By fitting the equation of state of lattice QCD at zero density, including entropy density $s/T^3$, pressure $P/T^4$, energy density $\epsilon/T^4$, trace anomaly $I/T^4$, specific heat $C_V$, and squared speed of sound $c_s^2$, the three parameters $c_1$ to $c_3$ in $V(\phi)$ can be determined. Then, by fitting the data of the second order baryon number susceptibility of lattice QCD at zero density $\chi_2^B$, the parameters $c_4$ and $c_5$ in $Z(\phi)$ can be fixed. 
As exhibited in Figure \ref{fig1}, the fitting outcomes reveal a notable alignment between our holographic model's predictions and the lattice QCD data. Notably, across all data points, the results obtained from our holographic model lie within the error bands of the lattice QCD data. The parameters employed in this fitting procedure are detailed in Table \ref{tab1}.
\begin{table}[htbp]
	\centering
	\begin{tabular}{|c|c|c|c|c|c|c|c|}
		\hline
		$c_1$  & $c_2$ & $c_3$ & $c_4$ & $c_5$ & $\kappa_5^2$ & $b$ & $\phi_s$ [MeV]\\
        \hline
		 $0.71$ & $3.7\times 10^{-3}$ & $2.8\times 10^{-5}$ & $0.35$ &  $0.0925$ & $3.36\pi$ & $-0.27435$ & $1085$\\
        \hline
	\end{tabular}
\caption{Parameters in Eq.~\eqref{eq51} by fitting the equation of state and second order baryon number susceptibility data from lattice QCD \cite{HotQCD:2012fhj, HotQCD:2014kol, Bazavov:2017dus}. }
\label{tab1}
\end{table}

\begin{figure}[htbp]
\centering
\includegraphics[width=.45\textwidth]{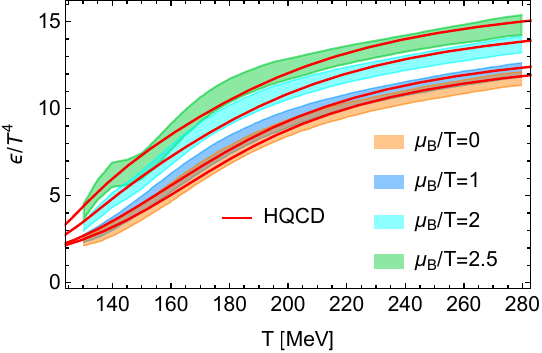}
\qquad
\includegraphics[width=.45\textwidth]{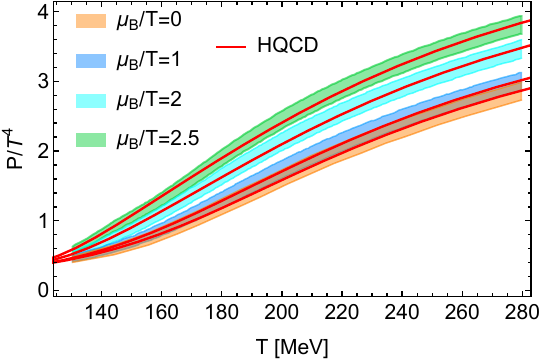}
\caption{The temperature dependence of $P/T^4$ (Left panel) and $\epsilon/T^4$ (Right panel) along lines with different $\mu_B/T$ ratios compared with lattice QCD data \cite{Bazavov:2017dus}. The bands show lattice QCD results, with red solid lines representing holographic QCD predictions. For $\mu_B/T= 0, 1, 2,~\text{and}~2.5$ , the holographic model aligns well with lattice QCD between $T= 130~\sim~ 280$ MeV.}\label{fig2}
\end{figure}
To further evaluate the quantitative accuracy of our holographic model, we analyzed the temperature evolution of $P/T^4$ and $\epsilon/T^4$ along lines with different values of the $\mu_B/T$, and compared it with lattice QCD data. As shown in Figure \ref{fig2}, within the temperature range of $130 \sim 280$ MeV, our holographic model's predictions are in good quantitative agreement with lattice QCD data across all four $\mu_B/T$ ratios ($0$, $1$, $2$, and $2.5$).

\begin{figure}[htbp]
\centering
\includegraphics[width=.32\textwidth]{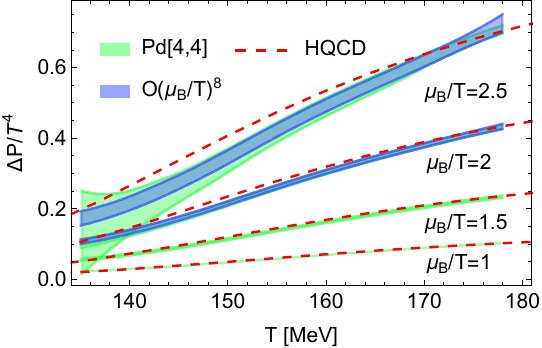}
\includegraphics[width=.32\textwidth]{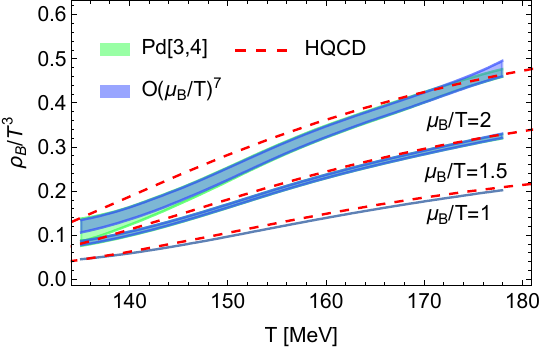}
\includegraphics[width=.32\textwidth]{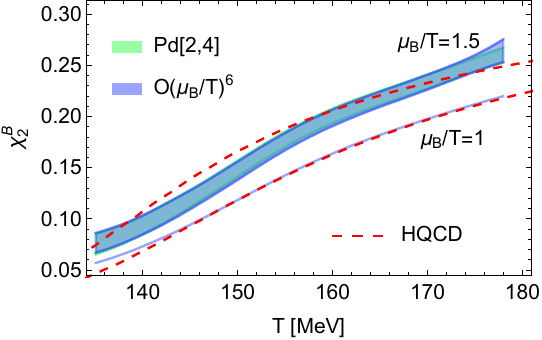}
\caption{A comparison of the temperature dependence of $\Delta P/T^4$  (Left panel) with $\Delta P(T,~\mu_B)=P(T,~\mu_B)-P(T,~0)$, $\rho_B/T^3$ (Middle panel), and 
 $\chi_2^B$ (Right panel) from holographic QCD with lattice QCD Taylor expansions and Pad$\acute{\text{e}}$ approximants data \cite{Bollweg:2022rps}. These comparisons are conducted along lines of varying $\mu_B/T$ ratios. The red dashed lines indicate predictions from holographic models, while the blue and green bands respectively correspond to lattice QCD Taylor expansions and Pad$\acute{\text{e}}$ approximants \cite{Bollweg:2022rps}. Different lines within each panel represent different  ratios.}\label{fig3}
\end{figure}
In Figure \ref{fig3}, we show the temperature dependence of three thermodynamic quantities: the pressure normalized to $\Delta P/T^4$ with $\Delta P(T,~\mu_B)=P(T,~\mu_B)-P(T,~0)$, the baryon density normalized to $\rho_B/T^3$, and the second order baryon number susceptibility $\chi_2^B$. The red dashed lines represent the predictions from holographic models. The blue and green bands correspond to the lattice QCD Taylor expansions and Pad$\acute{\text{e}}$ approximants \cite{Bollweg:2022rps}, respectively. Different lines within the same diagram represent varying $\mu_B/T$ ratios. The analysis is particularly focused on the temperature range of 135$\sim$180 MeV, which corresponds to the location of the QCD phase boundary. Within this temperature interval, all three thermodynamic quantities display a clear and consistent trend: a linear increase with rising temperature. A particularly notable observation is the variation in the slope of these linear increases as a function of the $\mu_B/T$ ratio. As $\mu_B/T$ increases, the slope of each quantity's temperature dependence becomes progressively steeper. This steepening of the slope implies that the phase transition grows more pronounced in intensity with higher $\mu_B/T$ ratios. This is a crucial finding as it offers a theoretical underpinning for the potential emergence of a first order phase transition at larger $\mu_B/T$ ratios.

\begin{figure}[htbp]
\centering
\includegraphics[width=.45\textwidth]{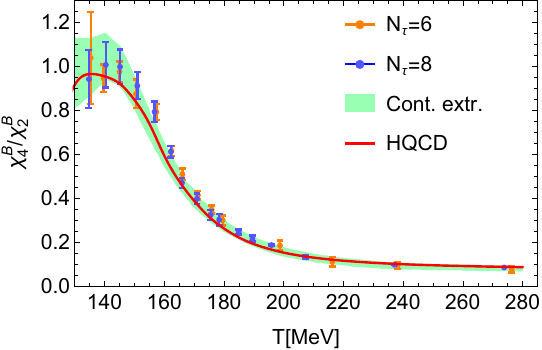}
\qquad
\includegraphics[width=.45\textwidth]{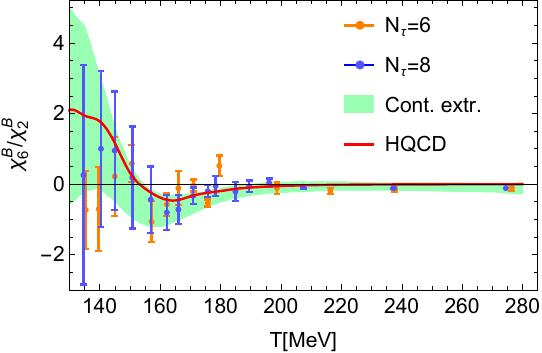}
\caption{Baryon number susceptibility $\chi^B_4/\chi^B_2$ (Left panel) and $\chi^B_6/\chi^B_2$ (Right panel) at $\mu_B=0$. The red solid line shows the holographic QCD results. Orange and blue data points with error bars respectively represent lattice QCD data for $N_\tau = 6$ and $N_\tau =8$ and the green band stands for the continuum extrapolated lattice QCD data \cite{Bazavov:2017dus, Bollweg:2022rps}.}\label{fig4}
\end{figure}
In Figure \ref{fig4}, we compare the holographic model results for higher order baryon number susceptibility ratios with lattice QCD data \cite{Bazavov:2017dus, Bollweg:2022rps}. The left panel shows the $\chi_4^B/\chi_4^B$ and the right panel shows the $\chi_6^B/\chi_4^B$. Near the phase boundary at $T\approx 156.5$ MeV, the $\chi_4^B/\chi_4^B$ peaks, while the $\chi_6^B/\chi_4^B$ shows a non monotonic trend, initially rising, then falling, and rising again. Notably, within the displayed temperature range, our holographic model results are almost indistinguishable from the previous holographic models \cite{Li:2023mpv}, with consistency observed even at the quantitative level. This indicates that whether $Z(0)=1$ has minimal impact on the behavior of baryon number susceptibilities near the phase boundary.

\begin{figure}[htbp]
\centering
\includegraphics[width=.32\textwidth]{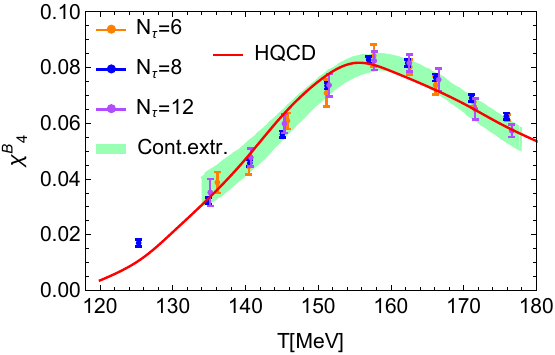}
\includegraphics[width=.32\textwidth]{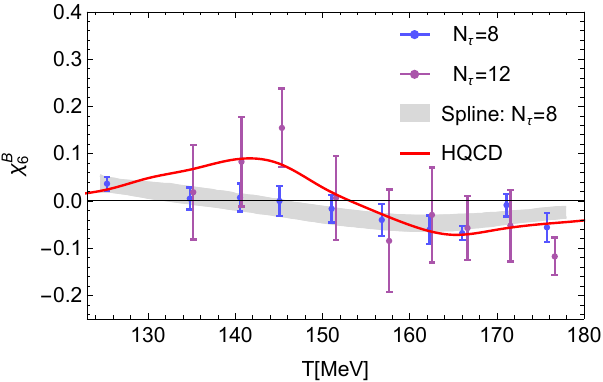}
\includegraphics[width=.32\textwidth]{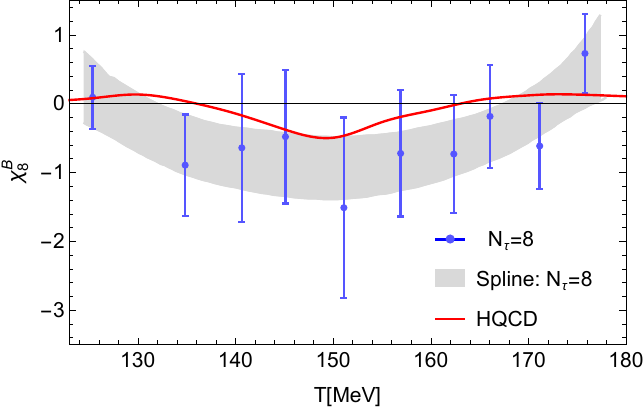}
\caption{Comparisons of holographic results (red solid lines) and lattice QCD data. From left to right are the $\chi^B_4$, $\chi^B_6$ and $\chi^B_8$, respectively. Green bands are lattice continuum extrapolated results \cite{Bollweg:2022rps}. Gray bands are lattice spline results \cite{Bollweg:2022rps}.}\label{fig5}
\end{figure}
In Figure \ref{fig5}, we compare our results for higher order baryon number susceptibilities with the latest lattice QCD data \cite{Bollweg:2022rps}. These results are focused on the region near the phase boundary, within the temperature range of 125$\sim$180 MeV. The panels from left to right respectively show the results for the $\chi_4^B$, $\chi_6^B$ and $\chi_8^B$. For the $\chi_4^B$, our results are in quantitative agreement with the lattice QCD continuum extrapolated results. Both show a rise and then a fall within this temperature range, peaking between 150$\sim$160 MeV. For the $\chi_6^B$, our holographic QCD results agree quantitatively with the $N_\tau = 12$ lattice QCD data. Both exhibit a non-monotonic trend: an initial increase, followed by a decrease, and then a final increase. They have a maximum between $T=140\sim 150$ MeV and a minimum between $T=155\sim 180$ MeV. The $\chi_8^B$ shows a more complex, non-monotonic temperature dependence. Its results are quantitatively consistent with previous holographic QCD studies \cite{Li:2023mpv}. This further confirms that the value of $Z(0)$ has little effect on the behavior of baryon number susceptibilities near the phase boundary.

\begin{figure}[htbp]
\centering
\includegraphics[width=.45\textwidth]{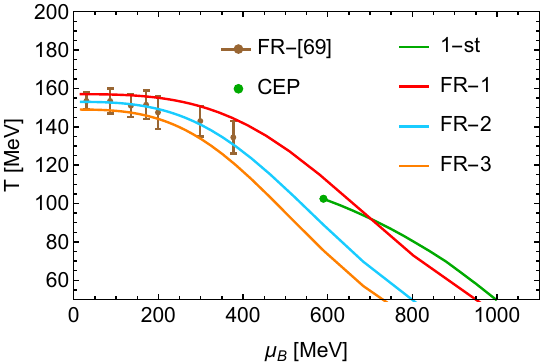}
\qquad
\includegraphics[width=.45\textwidth]{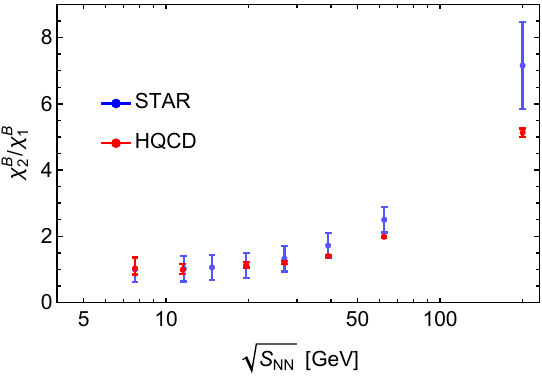}\\
\includegraphics[width=.45\textwidth]{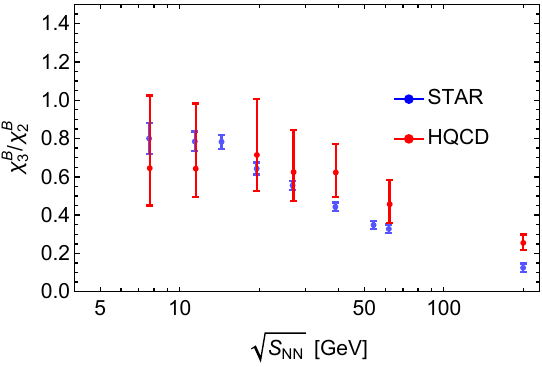}
\qquad
\includegraphics[width=.45\textwidth]{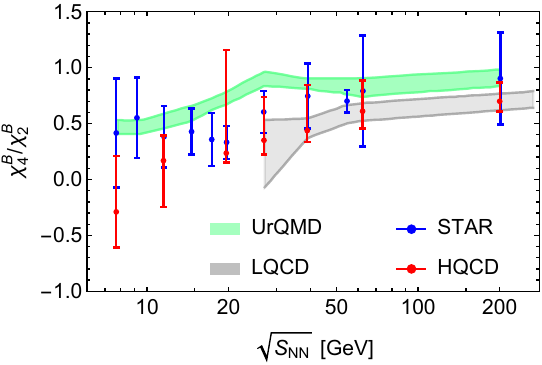}
\caption{The phase diagram (Top-left panel), $\chi_2^B/\chi_1^B$ (Top-right panel), $\chi_3^B/\chi_2^B$ (Bottom-left panel) and $\chi_4^B/\chi_2^B$ (Bottom-right panel) in holographic QCD model. The green dot represents the CEP ($T=102.25 $ MeV, $\mu_B=590.5$ MeV) and the green line stands for the first order phase transition line determined by the minimum free energy density. The brown data with error bars correspond to collision energies of $7.7\sim 200$ GeV from \cite{Gupta:2020pjd}. The red, cyan, and orange lines indicate the three fitted chemical freeze-out lines, labeled as FR-1, FR-2 and FR-3. Blue data points with error bars indicate the STAR Collaboration's net-proton distribution results for $0 - 5\%$ centrality Au-Au collisions \cite{Xu:2016mqs,STAR:2021fge,STAR:2025zdq}. Lattice QCD \cite{Bazavov:2020bjn,Bollweg:2024epj} and UrQMD results \cite{Bass:1998ca,Bleicher:1999xi} are represented as gray and green bands, respectively. Red data points with error bars show numerical results from holographic QCD.}\label{fig6}
\end{figure}
By comparing the equations of state and baryon number susceptibilities from our holographic model with lattice QCD data, we've found quantitative agreement in both zero and finite baryon density regions. This supports our model's predictive power at higher baryon densities. Consequently, we've calculated ratios of various baryon number susceptibilities in the finite baryon density region. To directly compare with experimental data, we've established relationships between collision energy $ \sqrt{S_{NN}}$ and $(T,~\mu_B)$ in two ways. First, we used the mappings provided in \cite{Gupta:2020pjd}, as shown in the top-left panel of Figure \ref{fig6}, where the brown data with error bars correspond to collision energies of $7.7\sim 200$ GeV from \cite{Gupta:2020pjd}. 

Using the correspondence between collision energies and positions on the phase diagram, we calculated the behavior of various ratios of baryon number susceptibilities as functions of collision energy as shown in Figure \ref{fig6}. The top-right, bottom-left, and bottom-right panels of Figure \ref{fig6} show the results of $\chi_2^B/\chi_1^B$, $\chi_3^B/\chi_2^B$ and $\chi_4^B/\chi_2^B$, respectively. Blue data points with error bars represent results from the STAR Collaboration for net-proton distributions for centrality $0 - 5\%$ Au-Au collisions \cite{Xu:2016mqs,STAR:2021fge,STAR:2025zdq}. Red data points with error bars are numerical results from holographic QCD. The centers correspond to the centers of the brown data points in the top-left panel, with errors derived from the errors of the brown data. The bottom-right panel also shows lattice QCD \cite{Bazavov:2020bjn,Bollweg:2024epj} and UrQMD results \cite{Bass:1998ca,Bleicher:1999xi} as gray and green bands, respectively. The holographic results aligns well with STAR data and lattice QCD and UrQMD results within uncertainties. And across the 7.7 - 200 GeV collision energy range, the $\chi_4^B/\chi_2^B$ exhibits a slow decrease at lower collision energies. Especially the central values of the holographic QCD results show quantitative agreement with lattice QCD \cite{Bazavov:2020bjn,Bollweg:2024epj}.

\begin{figure}[htbp]
\centering
\includegraphics[width=.45\textwidth]{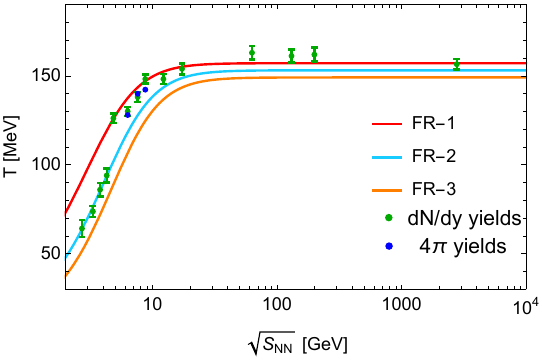}
\qquad
\includegraphics[width=.45\textwidth]{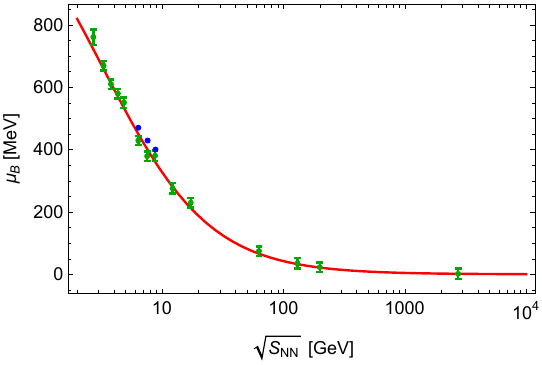}
\caption{The red, cyan, and orange lines indicate the three fitted chemical freeze-out lines. three representative chemical freeze-out lines using the $T-\sqrt{S_{NN}}$ and $\mu_B-\sqrt{S_{NN}}$ relationships in \cite{Andronic:2009jd, Andronic:2017pug} }\label{fig7}
\end{figure}
Second, we fitted three representative chemical freeze-out lines using the $T-\sqrt{S_{NN}}$ and $\mu_B-\sqrt{S_{NN}}$ relationships in \cite{Andronic:2009jd, Andronic:2017pug}. The fitting formula is as follows
\begin{equation}\label{eq52}
    T=\frac{T_0}{1+d_1 \left( 2+\sqrt{S_{NN}} \right)^{d_2}},~~\mu_B=\frac{1300}{1+0.3 \sqrt{S_{NN}}} ~~[\text{MeV}].
\end{equation}
Parameters for the three chemical freeze-out lines are set as shown in Table \ref{tab2}.
\begin{table}[htbp]
	\centering
	\begin{tabular}{|c|c|c|c|}
		\hline
	     & $T_0$ [MeV] & $d_1$ & $d_2$ \\
        \hline
	FR-1	& 157 & 60 & -2.85 \\       
        \hline
	FR-2	& 153 & 100 & -2.75 \\
         \hline
	FR-3	& 149 & 135 & -2.75 \\
        \hline
	\end{tabular}
\caption{Parameters for the three chemical freeze-out lines. }
\label{tab2}
\end{table}
The comparison between the fitted chemical freeze-out lines and the data in \cite{Andronic:2009jd, Andronic:2017pug} are shown in Figure \ref{fig7}. It is clearly that the second chemical freeze-out line which is labeled by "FR-2" has the best fit to the data. In the top-left panel of Figure \ref{fig6}, we show the positions of the three chemical freeze-out lines in the phase diagram. FR-1 has a higher temperature and passes above the CEP. FR-2 shows the best agreement with data in \cite{Gupta:2020pjd} and passes below and to the left of the CEP. FR-3 has the lowest temperature, also passes below and to the left of the CEP, but is further away from it than FR-2. These three curves help us understand how the ratios of baryon number susceptibility behave when passing through different regions and assist in analyzing potential CEP signals.

\begin{figure}[htbp]
\centering
\includegraphics[width=.45\textwidth]{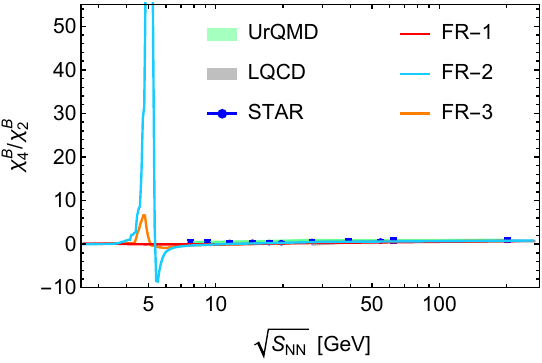}
\qquad
\includegraphics[width=.45\textwidth]{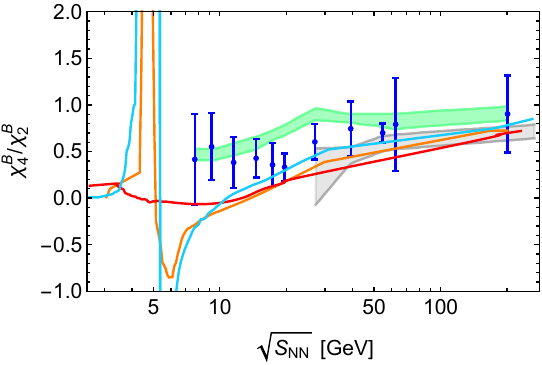}
\caption{Behavior of $\chi_4^B/\chi_2^B$ with collision energy $\sqrt{S_{NN}}$ along the three fitted chemical freeze-out lines. Blue data points with error bars represent STAR data \cite{Xu:2016mqs}. Gray and green bands represent lattice QCD \cite{Bazavov:2020bjn,Bollweg:2024epj} and UrQMD results \cite{Bass:1998ca,Bleicher:1999xi}, respectively.}\label{fig8}
\end{figure}
In Figure \ref{fig8}, we show how $\chi_4^B/\chi_2^B$ varies with collision energy $\sqrt{S_{NN}}$ along the three fitted chemical freeze-out lines. Blue data points with error bars represent STAR data \cite{Xu:2016mqs}. Gray and green bands represent lattice QCD \cite{Bazavov:2020bjn,Bollweg:2024epj} and UrQMD results \cite{Bass:1998ca,Bleicher:1999xi}, respectively. In the left panel, along FR-2 and FR-3, a non-monotonic collision energy dependence of $\chi_4^B/\chi_2^B$ is observed around 5 GeV. This corresponds to passing near the CEP in the phase diagram with $\mu_B\approx 520$ MeV. Here, as collision energy decreases, the ratio first forms a valley-like structure and then immediately a peak-like one. These features are more pronounced when the chemical freeze-out line passes closer to the CEP. The right panel shows that as collision energy decreases, the ratio $\chi_4^B/\chi_2^B$ along all three curves declines, reaching a minimum around 6 GeV, then quickly increases along FR-2 and FR-3 to form a peak at around 3$\sim$5 GeV before falling towards zero. Along FR-1, after the minimum around 6 GeV, the ratio gradually rises towards zero without forming a peak.

\section{Conclusions}\label{sec:06}

In this study, we present a refined holographic EMD model and analyze the critical behavior of phase transitions in hot and dense QCD matter within this framework. Our results indicate that, based on current experimental data, there is a high probability of observing a peak in $\kappa\sigma^2$ within the collision energy range of 3$\sim$5 GeV.

We began by outlining the general setup of the EMD model and employed holographic renormalization to calculate thermodynamic quantities. The action of the EMD model and its equations of motion were derived. To solve these equations, we specified the UV and IR boundary conditions through asymptotic expansions. Using the holographic dictionary, we determined the temperature and entropy density of the boundary finite-temperature field theory by calculating the black hole temperature and entropy. We then utilized holographic renormalization to regularize the EMD model action, obtaining a finite on-shell action. This allowed us to compute additional thermodynamic quantities such as free energy density, pressure, energy density, and trace anomaly via the boundary energy-momentum tensor.

We derived the first law of thermodynamics for a 5-dimensional bulk black hole spacetime using the Wald formalism and established a correspondence between bulk and boundary thermodynamic quantities. This confirmed that our holographic EMD model consistently describes the grand canonical ensemble of the boundary finite-temperature field theory.

Furthermore, we investigated the gravitational corrections to shear viscosity due to the Gauss-Bonnet term. By introducing a reasonable first-order background metric perturbation, we computed the n-2 form derived from the Wald formalism. We found that this n-2 form can be used to calculate the shear viscosity coefficient of the boundary fluid, and our results align with those in \cite{Charmousis:2012dw,Brigante:2007nu}.

By fitting lattice QCD equation of state data at zero density, we refined an EMD model to study critical behavior of QCD matter phase transitions. Our holographic model aligned with lattice QCD equation of state and baryon number susceptibility results within errors, even at finite density. Moreover, analyzing the baryon number susceptibility ratios across collision energies indicates that $\kappa\sigma^2$ is expected to exhibit a peak-like structure in the 3$\sim$5 GeV range.

\section{acknowledgments}
We are grateful to Yong Cai, Song He and Li Li for valuable discussions. This work was supported by the National Natural Science Foundation of China (Grant Nos. 12075213 and 12335005), the Natural Science Foundation for Distinguished Young Scholars of Henan Province (Grant No. 242300421046), the National Key Program for Science and Technology Research Development (2023YFB3002500), and the Natural Science Foundation of Henan Province of China (Grant No. 242300420235).

\begin{appendix}
\section{Wald method}
Here we give a brief review of the Wald method. We begin with the action
\begin{equation}
    S=\frac{1}{2\kappa_N^2}\int L
\end{equation}
with the n-form 
\begin{equation}
    L=\sqrt{-g}\mathcal{L} dx^1\wedge dx^2\dots \wedge dx^{n}
\end{equation}
The variation of all fields represented by $\Psi$
\begin{equation}
    \delta L=\text{EoM.} \delta \Psi +d\mathbf{\Theta}_\delta
\end{equation}
where EoM. is the equation of motion and $d\mathbf{\Theta}_\delta$ is the boundary term and $\delta$ means the boundary is reduced by $\Psi\rightarrow \Psi+\delta \Psi$. If we consider the variation of $\Psi$ caused by $x^\mu\rightarrow x^\mu+\xi^\mu$ as $\hat{\delta}\Psi=\mathbf{L}_\xi \Psi$ where $\mathbf{L}_\xi$ means the Lie derivative. Then
\begin{equation}
    \hat{\delta}L=\mathbf{L}_{\xi}L=d\left( i_\xi L \right)+i_\xi dL=d\left( i_\xi L \right)
\end{equation}
Then define a (n-1)-form
\begin{equation}
    \mathbf{j}=\mathbf{\Theta}_\xi -i_\xi L
\end{equation}
where $\mathbf{\Theta}_\xi$ means the boundary term reduced by $\Psi\rightarrow \Psi+\mathbf{L}_\xi \Psi$. Then we have 
\begin{equation}
    d\mathbf{j}=d\left(\mathbf{\Theta}_\xi -i_\xi L\right)=\hat{\delta}L-\text{EoM.} \mathbf{L}_\xi \Psi -\hat{\delta}L=-\text{EoM.} \mathbf{L}_\xi \Psi=\left. 0\right|_{on-shell}
\end{equation}
where "EoM." refers to the part that becomes zero after substitution into the equations of motion. Such that for on shell solutions we have
\begin{equation}
    \mathbf{j}=d\mathbf{Q}
\end{equation}
where $\mathbf{Q}$ is a (n-2)-form. For a variation $\delta$ with $\delta \xi=0$ we have
\begin{equation}
    \delta \mathbf{j}=\delta \mathbf{\Theta}_\xi -i_\xi \delta L=\delta \mathbf{\Theta}_\xi-\mathbf{L}_\xi \mathbf{\Theta}_\delta +d\left(i_\xi \mathbf{\Theta}_\delta\right)
\end{equation}
here we used $\mathbf{L}_\xi \Lambda=i_\xi d\Lambda+d\left(i_\xi \Lambda\right)$. Then we have 
\begin{equation}
    \delta \mathbf{j}-d\left(i_\xi \mathbf{\Theta}_\delta\right)=\delta \mathbf{\Theta}_\xi -\mathbf{L}_\xi \mathbf{\Theta}_\delta=\mathbf{\omega}(\Psi,\delta\Psi,\mathbf{L}_\xi \Psi)
\end{equation}
where $\mathbf{\omega}(\Psi,\delta\Psi,\mathbf{L}_\xi \Psi)$ is the (n-1)-form symplectic current defined as \cite{Wald:1993nt,Iyer:1994ys}
\begin{equation}
    \mathbf{\omega}(\Psi,\delta_1\Psi,\delta_2\Psi)=\delta_1\mathbf{\Theta}_{\delta_2}-\delta_2\mathbf{\Theta}_{\delta_1}
\end{equation}
Then the Hamiltonian corresponding to evolution by $\xi$ should satisfy
\begin{equation}
    \delta H_{\xi}=\int_\mathcal{C}\mathbf{\omega}(\Psi,\delta\Psi,\mathbf{L}_\xi \Psi)
\end{equation}
If we further let $\delta \Psi$ satisfies the equations of motion then we have the relation $\mathbf{j}=d\mathbf{Q}$ and
\begin{equation}
    \delta H_{\xi}=\int_\mathcal{C}\delta\mathbf{j}-\int_\mathcal{C}d\left(i_\xi \mathbf{\Theta}_\delta\right)=\int_{\partial\mathcal{C}}\left(\delta\mathbf{Q}-i_\xi \mathbf{\Theta}_\delta\right)
\end{equation}
which is purely a surface term. Note that if the vector $\xi$ is a Killing vector with $\mathbf{L}_\xi \Psi=0$. Then the $\mathbf{\omega}(\Psi,\delta\Psi,\mathbf{L}_\xi \Psi)=\mathbf{\omega}(\Psi,\delta\Psi,0)=0$.
In order to provide the first law of thermodynamics for black holes, we choose the vector $\xi=\partial_t$ which is a time-like Killing vector. Then we have 
\begin{equation}
    \delta H_{\xi}=\int_{\partial\mathcal{C}}\left(\delta\mathbf{Q}-i_\xi \mathbf{\Theta}_\delta\right)=0
\end{equation}
For our case the bulk $\mathcal{C}$ is the entire space, with its boundaries being the black hole horizon and the infinite boundary. And both $\mathbf{j}$ and $\mathbf{\Theta}$ are only related to $r$, so the integration of three-dimensional space is trivial, so we have
 \begin{equation}
    \delta H_{\xi}=\int_{\partial\mathcal{C}}\left(\delta\mathbf{Q}-i_\xi \mathbf{\Theta}_\delta\right)=\delta H_{\xi}^{\infty}-\delta H_{\xi}^{h}=0
\end{equation}

\end{appendix}

\providecommand{\href}[2]{#2}\begingroup\raggedright\endgroup

\end{document}